\documentclass[
 reprint,
 amsmath,amssymb,
 aps, 
 superscriptaddress,
 longbibliography
]{revtex4-1}

\usepackage{eurosym}
\usepackage{amsmath}
\usepackage{graphicx}
\usepackage{dcolumn}
\usepackage{bm}
\usepackage{mathtools}
\usepackage{ulem}
\usepackage[dvipsnames]{xcolor}
\usepackage{setspace}

\begin{document}

\preprint{APS/123-QED}

\title{Magnon spin transport in the van der Waals antiferromagnet CrPS\textsubscript{4} for non-collinear and collinear magnetization}
\author{Dennis K. de Wal}\email{d.k.de.wal@rug.nl}
\author{Muhammad Zohaib}
\affiliation{Zernike Institute for Advanced Materials, University of Groningen,
Groningen, the Netherlands}
\author {Bart J. van Wees}\affiliation{Zernike Institute for Advanced Materials, University of Groningen,
Groningen, the Netherlands}
\date{\today }

\begin{abstract}
We investigate the injection, transport and detection of magnon spins in the van der Waals antiferromagnet chromium thiophosphate (CrPS\textsubscript{4}). We electrically and thermally inject magnon spins by platinum contacts and examine the non-local resistance as a function of in-plane magnetic field up to 12 Tesla. We observe a large non-local resistance from both the electrically and thermally excited magnon modes above the spin-flip field where CrPS\textsubscript{4} is in the collinear state. At 25 K for an in-plane field of ranging 5 - 12 T, we extract the magnon relaxation length $\lambda_m$ ranging $200-800$ nm and a typical magnon conductivity of $\sigma_m\approx1\times10^4$ Sm\textsuperscript{-1}, which is one order of magnitude smaller than in yttrium iron garnet (YIG) films at room temperature. Moreover, we find that $\sigma_m$ is almost zero for CrPS\textsubscript{4} in the non-collinear state. In addition the non-local spin Seebeck effect shows a complex behavior as a function of field. Our results open up the way to understanding the role of the antiferromagnetic magnon modes on spin injection into antiferromagnets and implementation of two-dimensional magnets for scalable magnonic circuits.
\end{abstract}

\maketitle


\section{Introduction}
Magnon spin transport has been extensively studied in insulating ferro- and ferrimagnets, by spin pumping\cite{tserkovnyak_enhanced_2002}, the spin Seebeck effect (SSE)\cite{uchida_spin_2010} and electrical injection and detection\cite{cornelissen_long-distance_2015}. Antiferromagnets possess several advantages over ferromagnets for spintronic applications, such as stability to external fields\cite{wadley_electrical_2016} and operation frequencies up to terahertz scale\cite{baierl_terahertz-driven_2016}. Long distance magnon transport has been demonstrated in YFeO\textsubscript{3}\cite{das_anisotropic_2022} and antiferromagnetic hematite\cite{lebrun_long-distance_2020}, as well as coherent control of magnon spin dynamics\cite{wimmer_observation_2020}. The discovery of insulating ferro- and antiferromagnetic van der Waals materials such as the chromium trihalides (CrX\textsubscript{3}, X = Cl, Br, I) and transition metal phosphates (MPS\textsubscript{3}, M = Fe, Mn, Ni, Co) allows for the study of magnon spin transport in the quasi two-dimensional (2D) limit. These materials can be isolated into monolayer or few-layer thicknesses and show diverse inter- and intralayer exchange couplings and magnetic anisotropies. The resulting rich spin textures make 2D antiferromagnetic van der Waals materials a promising platform for the study of spin wave properties and transport.

\begin{figure}[t!]
\centering
\includegraphics[width=0.99\linewidth]{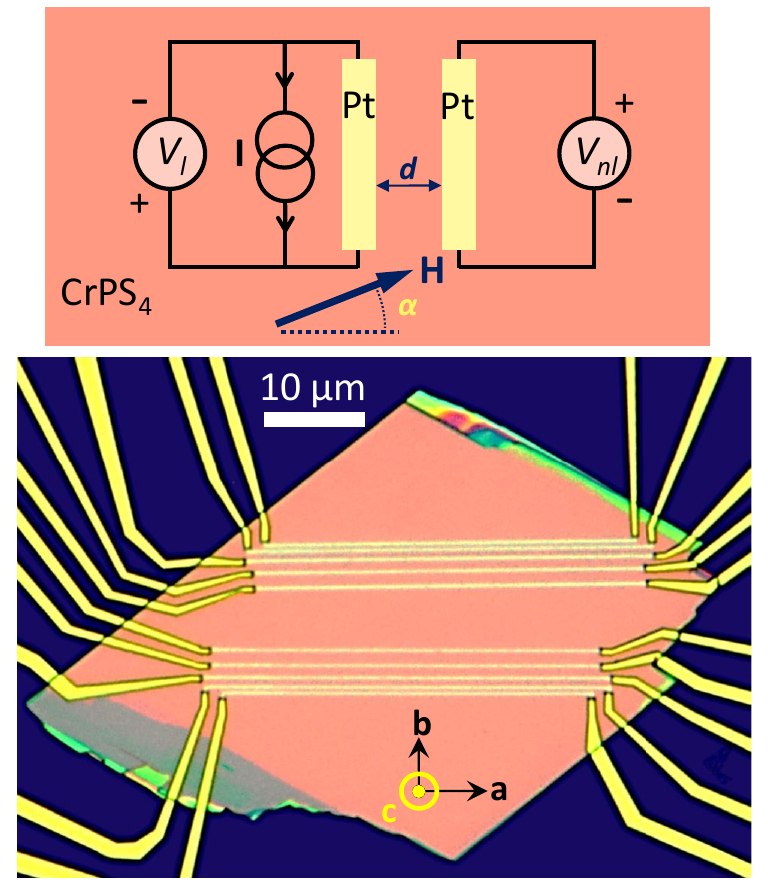}
\caption{Optical micrograph of a non-local magnon transport device with several parallel Pt strips bonded by Ti/Au leads on top of a $\sim$160 nm thick CrPS\textsubscript{4} exfoliated flake (crystal axes indicated). Top image shows the schematic electrical circuitry used in the non-local measurements. $\alpha$ is the in-plane magnetic field angle, normal (dashed line) to the Pt strips. $V_l$ and $V_{nl}$ are the voltages measured at the injector and detector Pt strip, respectively. }
\label{fig1:OpticalImage}%
\end{figure}

In these materials the investigation of antiferromagnetic resonance (AFMR) reveals the existence of acoustic and optical magnon modes \cite{macneill_gigahertz_2019,li_ultrastrong_2023}. However, AFMR studies are typically limited to sub-40 GHz excitation frequencies at external fields below 1.5 T, whereas the critical fields and precession frequencies of antiferromagnets often exceed these values. Moreover, AFMR does not resolve the role of the magnon modes in spin transport. The study of the propagation of magnons in van der Waals antiferromagnets, by employing a non-local geometry unveils information about the transport properties such as the magnon relaxation and magnon conductivity\cite{de_wal_long-distance_2023, brehm_magnon_2024} as well as the role of the antiferromagnetic magnon modes on the transport. 
\begin{figure*}[t!]
\centering
\includegraphics[width=0.99\textwidth]{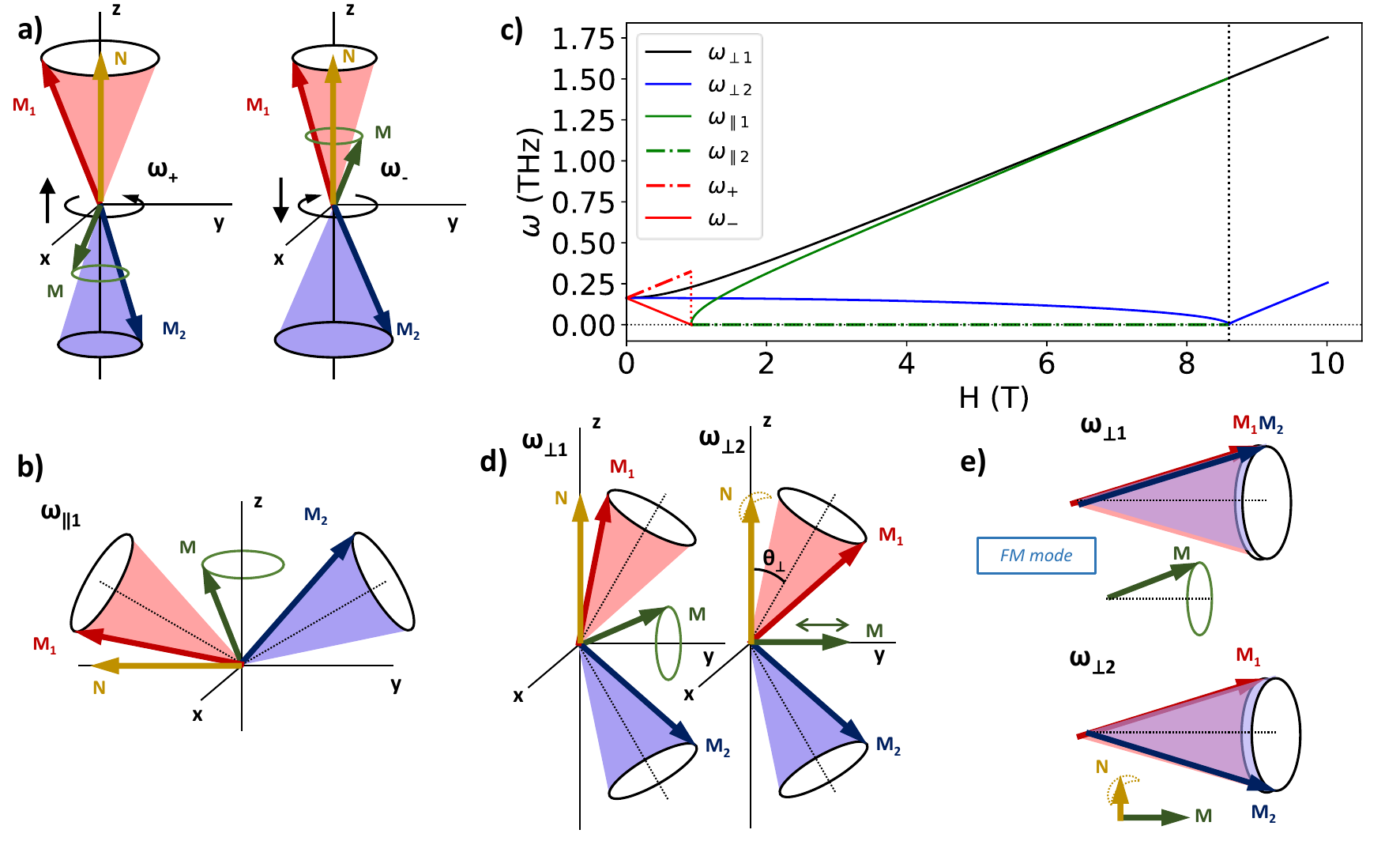}
\caption{Antiferromagnetic magnon modes in CrPS\textsubscript{4} for (a, b) an out-of-plane (oop) magnetic field in the z-direction, along $c$-axis, and (d, e) in-plane (ip) field in the y-direction, along $b$-axis. oop: (a) For $0<H<H_{sf}$ the $\omega_+$ and $\omega_-$ modes are non-degenerate in energy, each carrying opposite magnon spin in the z-direction. (b) $H_{sf}<H<H_{E\parallel}$ the spin flopped magnon mode ($\omega_{\parallel1}$) where the net magnetization $\mathbf{M}$ (green arrow)  precesses around $H$ and the Néel vector $\mathbf{N}$ (yellow arrow) changes only in magnitude in the y-direction. ip: (d) and (e): For $0<H<H_{E\perp}$, where the system is in the non-collinear state, the acoustic $\omega_{\perp1}$ and optical magnon mode $\omega_{\perp2}$ are depicted. For $\omega_{\perp1}$, $\mathbf{M}$ precesses around the ip external field (y-axis) and $\mathbf{N}$ (z-axis) changes only in magnitude (linearly polarized). For $\omega_{\perp2}$, $\mathbf{M}$ is linearly polarized in the direction of the field and $\mathbf{N}$ oscillates in the xz-plane and also changes its magnitude.
(e): For $H>H_{E\perp}$, collinear state, the acoustic mode is the ferromagnetic mode (Néel vector is zero) and the optical mode is the antiferromagnetic mode. (c) The dispersion relation of the eigenfrequencies for the different magnon modes at 2K is given for an anisotropy field $H_{ani}=0.1$ T (z-direction) and an exchange field $H_{exc}=4.25$ T\cite{peng_magnetic_2020, gurevich_magnetization_1996}. Figure adapted from:\cite{de_wal_long-distance_2023}}%
\label{fig2:Magnon_modes}%
\end{figure*}
Magnon transport driven by a thermal gradient (SSE) has been reported in both ferro- and antiferromagnetic van der Waals materials\cite{xing_magnon_2019, feringa_spin_2023,qi_giant_2023}. Yet these thermally driven magnons provide only convoluted information about the magnon transport properties. 

On the other hand, ``all electrical'' magnon transport does not suffer these problems, as the exact locations of magnon exitation and detection are clear, allowing for direct electrical control over the magnon currents. This enables the study of the spins carried by the different AFM magnon modes and their effect on transport. Yet, so far in antiferromagnetic van der Waals materials, field tunable all electrical long distance magnon transport has only been shown in CrPS\textsubscript{4} using angular dependent magnetoresistance (ADMR) measurements\cite{de_wal_long-distance_2023}. 

Here, we expand this work and perform a detailed study of the injection, transport and detection of spins by exciting the antiferromagnetic magnon modes for magnetic field perpendicular to the anisotropy axis ($c$-axis) of CrPS\textsubscript{4} through investigating the non-local resistance as a function of field and temperature. 

\section{Experimental details}
In this work, we employ a non-local geometry of 7 nm thick platinum strips on top of a $\sim$160 nm thick exfoliated CrPS\textsubscript{4} flake (Fig. \ref{fig1:OpticalImage}). A low frequency ($<20$Hz) AC charge current \textit{\textbf{I}} through an injector Pt strip generates an in-plane spin accumulation perpendicular to \textit{\textbf{I}} via the spin Hall effect (SHE)\cite{cornelissen_long-distance_2015}. Moreover, in the same Pt strip the current \textit{\textbf{I}} via by Joule heating, creates both vertical and lateral thermal gradients, in the CrPS\textsubscript{4} flake, which drives a spin Seebeck generated magnon current. Lastly, in an adjacent detector Pt strip, the two effects both generate non-local voltages V$_{el}$ and V$_{th}$ via the inverse SHE. The first comes from the injected spin current and the second from the spin Seebeck effect. 

The time-dependent voltage response for $I(t) = I_0 \sin{(\omega t)}$ can be expanded as $V(t)=R_1I(t) + R_2I^2(t) + ...$ where $R_1$ and $R_2$ are the first and second order resistances. The non-local resistance $R_{nl}=V$\textsubscript{detector}/$I$\textsubscript{injector} can therefore be separated as $R_{nl}^{1\omega}=V_{el}/I$ and $R_{nl}^{2\omega}=V_{th}/I^2$. Lebrun \textit{et al.} \cite{lebrun_tunable_2018} demonstrate that for the easy axis antiferromagnet hematite ($\alpha$-Fe\textsubscript{2}O\textsubscript{3}), at fields below the spin-flop transition, the magnonic spin currents carry spin polarized along the Néel vector $\mathbf{N} = (\mathbf{m_1} -\mathbf{m_2})/2$, where $\mathbf{m_1}$ and $\mathbf{m_2}$ are the sublattice magnetizations. Yet, they show that thermal spin currents scale with the net (field-induced) magnetic moment $\mathbf{M} = (\mathbf{m_1} + \mathbf{m_2})/2$. However, the effect of the magnon modes themselves on the transport is not discussed.

In figure \ref{fig2:Magnon_modes}, the (k=0) magnon modes for CrPS\textsubscript{4} are depicted for non-zero out-of-plane (oop) and in-plane (ip) external field. Figure \ref{fig2:Magnon_modes}a and \ref{fig2:Magnon_modes}b show the modes before and after the spin-flop transition (oop), respectively. 
For the $\omega_{\parallel1}$ mode (acoustic) the Néel vector is in the y-direction (Fig. \ref{fig2:Magnon_modes}b). Under the assumption that the ip anisotropies $H_a=H_b$ (where the subscript indicated the crystal axis), the frequency of $\omega_{\parallel2}$ (optical) vanishes at the spin-flop. We expect that $\omega_{\parallel2}$ becomes a soft magnon mode, with zero frequency. In this case only the acoustic mode $\omega_{\parallel1}$ has a non-zero frequency\cite{li_ultrastrong_2023}. Figure \ref{fig2:Magnon_modes}d and \ref{fig2:Magnon_modes}e show the non-collinear (canted AFM) and the collinear (FM) state (for ip), before and after the spin-flip field above which the two sublattices align with the field, respectively. For ip fields, the ip projection (y-direction) of the net magnetization $\mathbf{M}$ increases linearly with increasing field till the spin-flip field ($H_{E\perp}=2H_{exc}+H_{ani}$)\cite{de_wal_long-distance_2023, gurevich_magnetization_1996}. The Néel vector $\mathbf{N}$ is along the z-axis and decreases with increasing field. 
The canting angle $\sin{\theta_{\perp}} = H/H_{E\perp}$ is the angle between the static sublattice magnetization and the anisotropy axis (z-axis). 

At $H>H_{E\perp}$ the system is in the collinear state. The acoustic and optical modes persist; the first, $\omega_{\perp1}$, is a FM-mode, which is the same as for an uniaxial ferromagnet (also known as Kittel mode).
The latter, $\omega_{\perp2}$, is the AFM-mode, 
where $\mathbf{M}$ is static and $\mathbf{N}$ still oscillates in the xz-plane. The dispersion relation at the band edge (k=0) as a function of the external field (at T=2 K) is shown in figure \ref{fig2:Magnon_modes}c for both the in-plane and out-of-plane magnon modes. 

A recent study on spin pumping in CrCl\textsubscript{3} reveals that in the spin-flopped state the acoustic mode ($\omega_{\parallel1}$) the spin current is driven by $\mathbf{M}$, yet for the optical mode ($\omega_{\parallel2}$) it is driven by $\mathbf{N}$\cite{wang_nevector_2022}. Further, a recent theoretical study by P. Tang \textit{et al.}\cite{tang_thermal_2024} on spin pumping in the non-collinear AFMs shows that when the two modes contribute equally, the spin current follows $\mathbf{M}$, whereas the spin current component along $\mathbf{N}$ only plays a role when the modes do not contribute equally (uncompensated Pt/AFM interface). 
Despite this understanding, the effect of the antiferromagnetic magnon modes on the transport of magnons, especially in the non-collinear regime with field perpendicular to the anisotropy axis, remains unclear.

Therefore, we explore here which antiferromagnetic magnon modes can be excited and detected by a spin accumulation $\boldsymbol{\mu}$. In the Pt strips, at the Pt/CrPS\textsubscript{4} interface, produced by the SHE, the spin accumulation $\boldsymbol{\mu}$ can only be generated in y-direction. 
Hence, when a magnon mode can carry spin in the y-direction it can absorb $\boldsymbol{\mu}$. For CrPS\textsubscript{4} where we apply field $H$, we can distinguish 5 phases: For oop fields: 1. $H<H_{sf}$ (where $H_{sf}=\sqrt{2H_{exc}H_{ani}}$ is the spin-flop field); spin injection is only possible for $H\parallel\boldsymbol{\mu}$ and $\boldsymbol{\mu}\parallel M,N$, 2. $H_{sf}<H<H_{E\parallel}$ \& 3. $H>H_{E\parallel}$ ($H_{\parallel}=2H_{exc}-H_{ani}$ is the spin-flip field parallel to the anisotropy axis). For ip fields: 4. $0<H<H_{E\perp}$ \& 5. $H>{E\perp}$; spin injection is only possible for $\boldsymbol{\mu}\parallel H$ and $H\parallel M$, $H\perp N$. 
Secondly, only the magnon modes in which $\mathbf{M}$ or $\mathbf{N}$ precess can pump spins into the Pt. Since $\boldsymbol{\mu}$ in the Pt is only collinear with $H$ in phase 4. and 5. and only  for the acoustic magnon mode $\omega_{\perp1}$ $\mathbf{M}$ precesses, only $\omega_{\perp1}$ contributes to spin pumping. By reciprocity, electrical injection of spin can therefore only excite $\omega_{\perp1}$.

Nonetheless, the above depiction only holds under the assumption that both magnetic sublattices contribute equally to the spin pumping (and injection). With CrPS\textsubscript{4} being an A-type antiferromagnet, the second sublattice has a greater distance to the Pt/CrPS\textsubscript{4} interface than the first. This could lead to a magnetically uncompensated interface
, for which precession of $\mathbf{M}$ or $\mathbf{N}$ orthogonal to the y-axis will contribute to the spin pumping as well\cite{tang_thermal_2024}. The two spin currents from CrPS\textsubscript{4} to the Pt as a result of spin pumping originating from the two individual sublattices carry spin collinear to their static sublattice magnetization (dashed line in figure \ref{fig2:Magnon_modes}d). The spin polarization of these spin currents over the interface have equal projection on the y-axis, but opposite on the z-axis. Hence, when both sublattices do not couple equally, the optical magnon mode $\omega_{\perp2}$ can contribute to the spin pumping (and injection) proportional to oscillating $\mathbf{N}$.

\begin{figure*}[ptb]
\centering
\includegraphics[width=\textwidth]{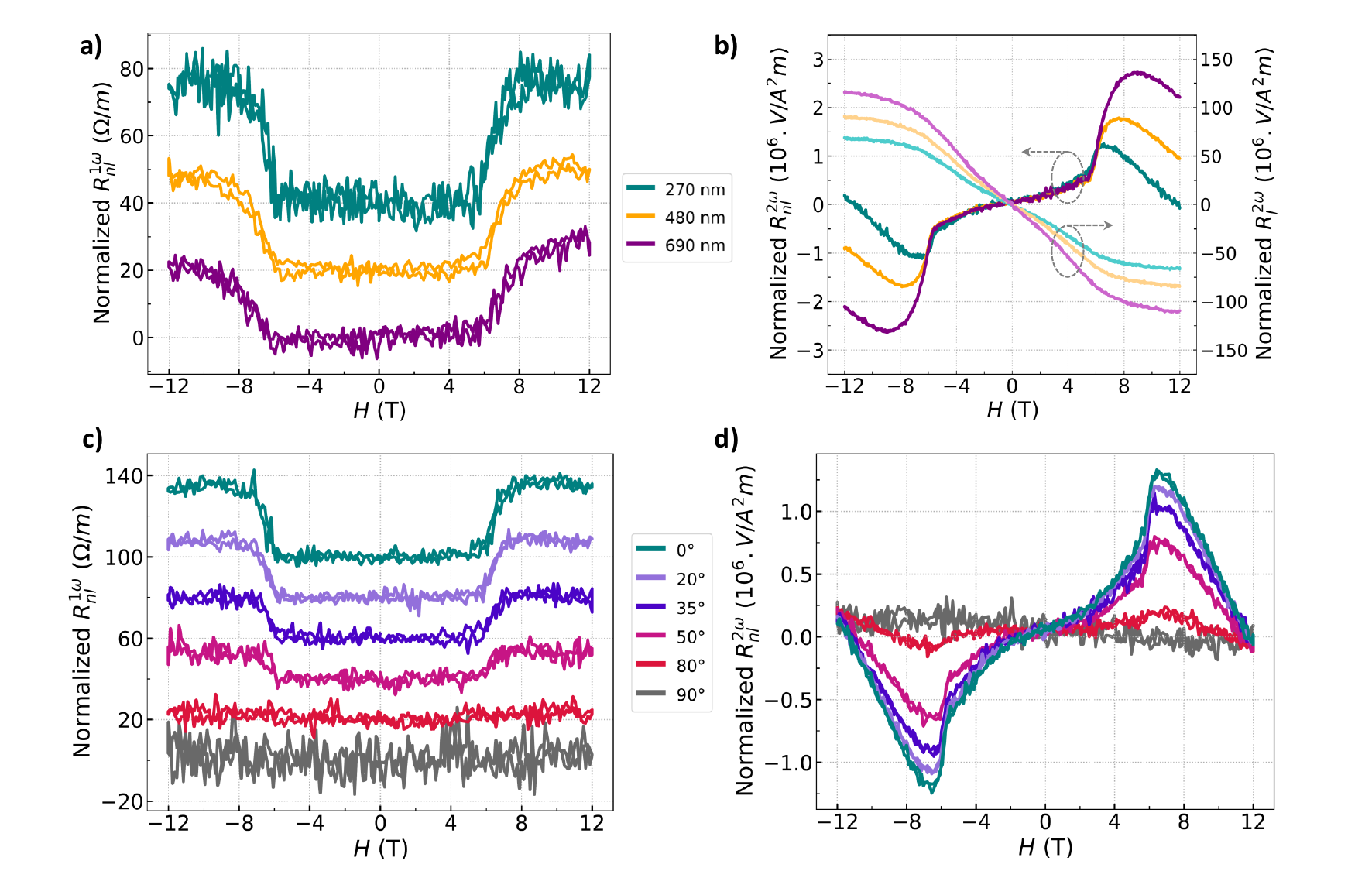}
\caption{Magnetic field dependence of $R_{nl}^{1\omega}$, $R_{nl}^{2\omega}$, and $R_{l}^{2\omega}$ at different $d$ and oop angle $\beta$ at 25 K for an AC driving current of 60$\mu$A. (a) and (b), $R_{nl}^{1\omega}$, $R_{nl}^{2\omega}$ (left axis) and $R_{l}^{2\omega}$ (right axis), respectively, as a function of field along the y-axis for different $d$. For $R_{nl}^{1\omega}$ in (a) a constant offset of -27.3 $\Omega/m$, -4.9 $\Omega/m$, and -6.6 $\Omega/m$ is removed, for $d=$270 nm, $d$=480 nm, and $d$=690 nm, respectively. The $R_{nl}^{1\omega}$ for $d=270$ nm and $d=480$ nm are shifted upwards for clarity by 40 $\Omega/m$ and 20 $\Omega/m$, respectively. For $R_{l}^{2\omega}$ a constant offset is removed as well. (c) and (d), $R_{nl}^{1\omega}$ and $R_{nl}^{2\omega}$, respectively, are plotted for different oop angle $\beta$ for an AC driving current of 80 $\mu$A, $d=$ 270 nm. Also in (c) a constant offset is removed and the curves are spaced by 20 $\Omega/m$ steps. In all plots the outliers are removed.}
\label{fig3:1omega}%
\end{figure*}

\section{\label{sec:data} Results and discussion}
In figure \ref{fig3:1omega}a, $R_{nl}^{1\omega}$ is plotted at 25 K as a function of magnetic field along the y-axis (ip). We find that at $H<$ 6 T, the system is in the non-collinear state where $R_{nl}^{1\omega}$ is zero. At 6 T $<H<$ 8 T, $R_{nl}^{1\omega}$ increases till it saturates at $H>$ 8 T. This behavior agrees with the magnetization of the CrPS\textsubscript{4} at 25 K, given in the Supporting Information (SI)\cite{Supplementary_Mat_2024},\cite{cornelissen_magnon_2016,takahashi_spin_2003, wei_giant_2022}
, where $\mathbf{M}$ for an in-plane field gradually saturates for 6 T $<H<$ 8 T\cite{peng_magnetic_2020}. $R_{nl}^{1\omega}$ in figure \ref{fig3:1omega} is given for three injector detector spacings $d$, which is the edge-edge distance between the Pt strips. These results agree well with the $R_{nl}^{1\omega}$ obtained from ADMR measurements in our earlier work\cite{de_wal_long-distance_2023}. 

At 25 K, $k_BT \gg \hbar\omega$, with $\omega$ being the frequency of the magnon modes, hence thermal equilibrium magnons populate both magnon modes (see figure \ref{fig2:Magnon_modes}c) at all field strengths shown in figure \ref{fig3:1omega}a. Therefore, both modes ($\omega_{\perp1}$ and $\omega_{\perp2}$) could contribute to transport. Regardless of which magnon mode contributes, the absence of $R_{nl}^{1\omega}$ below $H_{E\perp}$ is surprising. Possibly, the spin is not conserved when CrPS\textsubscript{4} is in the canted AFM state, due to the axial symmetry breaking\cite{tang_thermal_2024,bauer_soft_2023}. We discuss other possible reasons later. 

The thermally generated magnon transport signal shows a very different trend as a function of field. For $R_{nl}^{1\omega}$, either when there is a magnon depletion or accumulation, created by injection of spin generated by the SHE in the injector, depending on the sign of the SHE, the parallel or antiparallel configuration of the generated magnon spins w.r.t. M is compensated by the equal sign of the ISHE in the detector. This makes $R_{nl}^{1\omega}$ symmetric in field. The sign of $R_{nl}^{2\omega}$ does depend on the relative orientation of $\mathbf{M}$ w.r.t. the spin accumulation at the Pt/CrPS\textsubscript{4} interface in the detector contact. The latter makes $R_{nl}^{2\omega}$ antisymmetric in field\cite{cornelissen_long-distance_2015}. In figure \ref{fig3:1omega}b the local (right axis) and non-local (left axis) $R^{2\omega}$ are shown for different $d$. The advantage of the local over the non-local SSE signal is that the first contains direct information on effect of the magnon modes on the thermal spin pumping. Whereas the latter contains convoluted information on the magnon transport as well. For the local SSE ($R^{2\omega}_{l}$), in the non-collinear state, the number of thermally excited magnons increases with increasing $\sin{\theta_{\perp}}$, i.e. follows the net magnetization (ferromagnetic SSE)\cite{tang_thermal_2024}. However, at fields close to $H_{E\perp}$ $R^{2\omega}_{l}$ increases non-linearly w.r.t $\sin{\theta_{\perp}}$. Upon saturation of the sublattice magnetization, the signal starts to saturate, yet for $H>H_{E\perp}$ the saturation continues up to higher fields than for $R_{nl}^{1\omega}$ (Fig. \ref{fig3:1omega}a).

\begin{figure*}[ptb]
\centering
\includegraphics[width=0.99\textwidth]{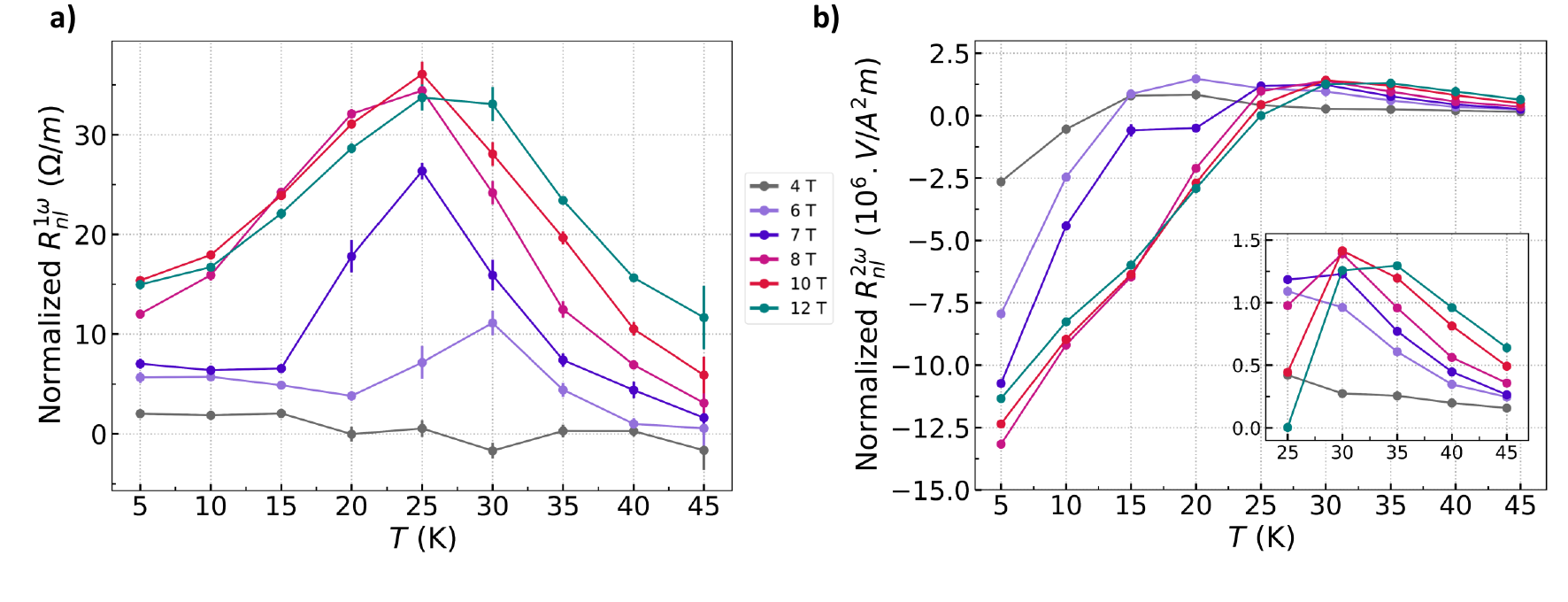}
\caption{The dependence of $R_{nl}^{1\omega}$ (a) and $R_{nl}^{2\omega}$ (b) on temperature for given fields at 100 $\mu$A of AC driving current for $d=270$ nm. In (b) the inset is a zoom in for temperatures 25 K - 45 K.}
\label{fig4:temp}%
\end{figure*}

For the non-local SSE signal $R_{nl}^{2\omega}$, at $H<H_{E\perp}$, (Fig. \ref{fig3:1omega}b), the dependence on the field is similar to that of the local SSE signal. However, the sign of $R_{nl}^{2\omega}$ is opposite to $R_{l}^{2\omega}$ and the amplitude is two orders of magnitude smaller. The former indicates that, driven by the SSE, the magnon chemical potential is opposite. The behavior of $R_{nl}^{2\omega}$, can be understood as follows: A thermal magnon current (proportional to $\nabla T$) is driven away from the injector by the SSE, effectively creating a depletion of magnons at the injector and a magnon accumulation away from the injector. This accumulation drives diffusive magnon currents, proportional to $\nabla\mu_m$,  towards the injector and detector\cite{cornelissen_nonlocal_2017}. For $H<H_{E\perp}$, the absence of a $R_{nl}^{1\omega}$ is this state suggests that diffusive transport lengths are very small and the similarity of $R_{nl}^{2\omega}$ to $R_{l}^{2\omega}$ and in this state points out that the SSE drives the magnons till just under the detector, where the magnon accumulation is detected. 

The increase at $H_{E\perp}$, which is very similar to that in $R_{nl}^{1\omega}$ (figure \ref{fig3:1omega}a), shows the onset of diffusive magnon transport. Yet, where $R_{nl}^{1\omega}$ clearly saturates at larger fields, $R_{nl}^{2\omega}$ decreases and even changes sign at fields far above $H_{E\perp}$ for the shortest $d$. This can be understood by the competing thermal (SSE-driven) and diffusive magnon current in the system, i.e. both the SSE generated magnon accumulation and diffusive magnon currents contribute to $R_{nl}^{2\omega}$. The decrease in $R_{nl}^{2\omega}$ at largest field strengths could indicate that the diffusion lengths increase with increasing field. To our surprise, for larger $d$, $R_{nl}^{2\omega}$ actually becomes larger. In YIG a sign change of $R_{nl}^{2\omega}$ is observed as a function of $d$. At larger $d$ (where $d\ll\lambda_m$ still holds), $R_{nl}^{2\omega}$ increases with increasing $d$\cite{cornelissen_nonlocal_2017}. Similar behavior could be possible in CrPS\textsubscript{4}.

In figure \ref{fig3:1omega}c, $R_{nl}^{1\omega}$ as a function of field at different oop angles ($\beta$) is shown. Note that $\beta=90^{\circ}-\theta_{\perp}$. Within the non-collinear regime $R_{nl}^{1\omega}$ remains zero, whereas in the collinear regime the $R_{nl}^{1\omega}$ scales with the projection of $\mathbf{M}$ on the y-axis ($\cos^2{\beta}$, as being dependent on the electron spin accumulation at both the injector and detector). For $R_{nl}^{2\omega}$, (figure \ref{fig3:1omega}d), the signal also scales with the projection of $\mathbf{M}$ ($\cos{\beta}$). For $\beta=90^\circ$, where the spin-flop is predicted around 0.8 T, neither for $R_{nl}^{1\omega}$, nor for $R_{nl}^{2\omega}$ this transition is observed. As we cannot see any contribution of oop spins and since $\mu\parallel\mathbf{N}$, following the spin pumping in CrCl\textsubscript{3}\cite{wang_nevector_2022}, only the optical mode ($\omega_{\parallel2}$) could contribute to spin pumping. However, this mode is a soft mode in our system. 

In figure \ref{fig4:temp}, $R_{nl}^{1\omega}$ and $R_{nl}^{2\omega}$ are given as a function of temperature. In the non-collinear regime $R_{nl}^{1\omega}$ is zero for all temperatures above and below the Néel temperature. In the collinear regime a non-monotonous dependence is observed. For temperatures $<$10 K (at H$>$7 T) the acoustic magnon mode ($\omega_{\perp1}$) is possibly not occupied, but $R_{nl}^{1\omega}$ is non-zero. This suggests that the optical magnon mode ($\omega_{\perp2}$) does contribute spin pumping, indicating an unequal coupling of the sublattices. Around 25 K, $R_{nl}^{1\omega}$ is maximum and at higher temperature $R_{nl}^{1\omega}$ decreases, diminishing just above the Néel temperature, T\textsubscript{N} = 38 K) (at larger field strenghts, the magnetic ordering is maintained up to slightly higher temperatures, see magnetization behavior in the SI). 
However, the number of magnons and the magnon transport properties are highly temperature dependent, therefore, information on the contributions by the different magnon modes cannot be directly extracted.

In addition, $R_{nl}^{2\omega}$ shows a very different behavior as a function of temperature. The measured $R_{nl}^{2\omega}$ changes sign for low temperature ($<$15 K) (see SI section 7), where the effect is strongest in the collinear regime. The local SSE $R_{l}^{2\omega}$, given in the SI, does not show this behavior. The occupation of the acoustic magnon mode ($\omega_{\perp1}$) is affected by the temperature, whereas the optical mode ($\omega_{\perp2}$) will remain populated for all temperatures in figure \ref{fig4:temp}b. For fields strength of $\leq$7 T, the eigen energy $\hbar\omega_{\perp1}<k_BT$, the acoustic mode could contribute spin pumping. However, the effect of the temperature on the non-local SSE in CrPS\textsubscript{4} and the effect of the magnon modes on transport are not fully understood. 


\begin{figure*}[ptb]
\centering
\includegraphics[width=0.99\textwidth]{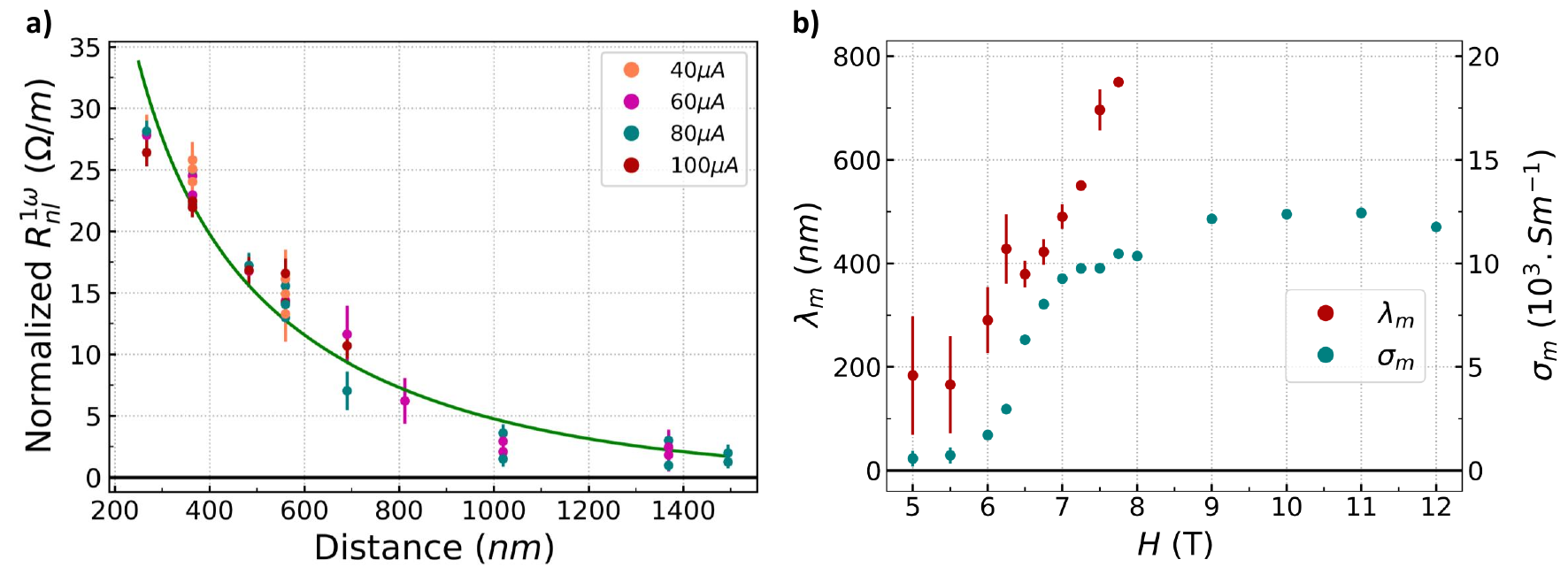}
\caption{The dependence of $R_{nl}^{1\omega}$ on the edge-edge distance $d$ between the injector and detector Pt contact. (a) Data obtained from three different samples, both from ADMR and field dependent (see figure \ref{fig3:1omega}a) measurements at different driving currents ranging from 40 $\mu$A to 100 $\mu$A at 25 K, 7 T. The green curve represents the best fit of equation \ref{eq:Magnon_decay}. The fitting is performed using least square minimization process where uncertainty of each data point is weighted according to the magnitude of the error. (b) $\lambda_m$ (in \color{Mahogany} \textbf{bordeau}\color{black}) and $\sigma_m$ (in \color{PineGreen}\textbf{teal}\color{black}) obtained by fitting the distance dependent $R_{nl}^{1\omega}$ at different fields around the spin-flip transition. For $H>8$T, $\sigma_m$ is estimated under the assumption $\lambda_m \gg d$, see SI}%
\label{fig5:distance}%
\end{figure*}

Altogether, the injection (excitation) of magnon spins and detection (incoherent spin pumping), and the transport of magnon spins by the magnon modes are entirely different processes. Spin is not a conserved quantity. This holds for injection and detection, and also for magnon spin transport. Thus, magnons excited by a spin current at the injector can diffuse towards the detector where they again create a spin current via spin pumping, but the spin is not necessarily conserved during transport (or injection and detection). Therefore, in our transport measurements we cannot determine if the magnon modes `carry' the spins. Furthermore, the magnon relaxation might be affected by the non-collinearity of the AFM. When the two sublattices are non-collinear, the strong exchange interaction possibly suppresses the magnon conductivity $\sigma_m$ and relaxation length $\lambda_m$. At the spin flip field, the exchange energy is overcome and the sublattices become collinear, possibly allowing a strong increase in both $\sigma_m$ and $\lambda_m$\cite{tang_thermal_2024,bauer_soft_2023}.  

In figure \ref{fig5:distance}a, the $R_{nl}^{1\omega}$ in the collinear state (at 7 T) is given as a function of edge-edge distance, $d$, between the injector and detector Pt strip, measured for multiple devices. The model for diffusive magnon transport leads to a decay in $R_{nl}^{1\omega}$ with increasing $d$ as a function of the $\lambda_m$. Under the assumption of a large enough effective interface spin mixing conductance, $R_{nl}^{1\omega}$ is given by:
\begin{equation}
    R_{nl}^{1\omega} = \frac{C}{\lambda_m}\frac{\exp{d/\lambda_m}}{1-\exp{2d/\lambda_m}},
    \label{eq:Magnon_decay}
\end{equation}
where C is a constant capturing all distance-independent pre-factors, such as the magnon conductivity $\sigma_m$. For $d<\lambda_m$ the transport is purely diffusive and $R_{nl}^{1\omega} \sim 1/d$, with the transport being Ohmic\cite{cornelissen_long-distance_2015}, whereas for $d>\lambda_m$, $R_{nl}^{1\omega}$ decays exponentially as the magnon relaxation sets in. From equation \ref{eq:Magnon_decay} we extract $\lambda^{1\omega}_m= 490\pm30$ nm with $\sigma_m\approx1\times10^4$ Sm$^{-1}$ , at 7 T, under the assumptions elaborated in SI.

In figure \ref{fig5:distance}b, $\lambda_m$ and $\sigma_m$ are extracted using equation \ref{eq:Magnon_decay}, in similar fashion as in Fig. \ref{fig5:distance}a, as functions of the ip field around the spin-flip transition. At 5 T, $\lambda_m$ is smaller than the smallest $d$ on our devices and $\sigma_m$ is (close to) zero. Here, the magnon transport is heavily suppressed. Increasing in field, between 6-8 T, both transport parameters increase. $\sigma_m$ increases sharply and saturates at field $>$8 T, whereas the increase in $\lambda_m$ is less abrupt and, in our measurements, does not seem to saturate. In fact, $>$8 T, the decay of $R_{nl}^{1\omega}$ as a function of $d$ seems to be fully Ohmic, yet due to insufficient data we cannot extract $\lambda_m$ at these fields (see SI). For these fields we extract $\sigma_m$ by assuming $\lambda_m \gg d$, see SI.

The effect of the different magnon modes on magnon spin transport is still not entirely disclosed. For systems using the non-local geometries the measured non-local resistance depends on several factors, the SHE and ISHE in the Pt contacts, the transparency of the Pt/CrPS\textsubscript{4} interface and the magnon conductivity and relaxation. 
The large spin Hall magnetoresistance measured on these samples (see SI) and in previous work\cite{de_wal_long-distance_2023} indicates a transparent interface.   

\section{Conclusion}
The exact effect of the antiferromagnetic magnon modes on magnon spin transport in the uniaxial antiferromagnet CrPS\textsubscript{4} with field orthogonal to the anisotropy axis, remains so far unclear. In the non-collinear regime the magnon relaxation length $\lambda_m$ and the magnon conductivity $\sigma_m$ are limited. We find that both $\lambda_m$ and $\sigma_m$ strongly increase at the spin-flip transition. At 7 T, we find $\lambda_m= 490\pm30$ nm and $\sigma_m\approx1\times10^4$ Sm$^{-1}$, the latter is one order of magnitude smaller than the typical values found in 210 nm thick YIG at room temperature\cite{cornelissen_long-distance_2015}. The thermally generated magnons via the SSE indicates that both magnon modes contribute to the non-local resistance, yet their individual contributions to transport continues to be unresolved. Obviously, currently we do not have a full comprehension of the role of the various modes for magnon spin transport. Nevertheless, these result prepare the way to understanding and using the antiferromagnetic magnon modes for long-distance magnon spin transport in 3D and 2D van der Waals antiferromagnets.  

\begin{acknowledgments}
We want to express our special gratitude towards G.E.W. Bauer, P. Tang and J. Barker for insightful discussions and suggestions. We acknowledge the technical support from J. G. Holstein, H. Adema, H. H. de Vries, and F. H. van der Velde. We acknowledge the financial support of the Zernike Institute for Advanced Materials and the European Union’s Horizon 2020 research and innovation program under Grant Agreements No. 785219 and No. 881603 (Graphene Flagship Core 2 and Core 3). This project is also financed by the NWO Spinoza prize awarded to B.J.W. by the NWO and has received funding from the European Research Council (ERC)
under the European Union’s 2DMAGSPIN (Grant Agreement No. 101053054). 
\end{acknowledgments}

\bibliography{Reference}

\pagebreak
\widetext






\setcounter{figure}{0}
\renewcommand{\thefigure}{S\arabic{figure}}
\renewcommand{\theequation}{S\arabic{equation}}
\setcounter{section}{0}
\renewcommand{\thesection}{S\arabic{section}}

\begin{center}
\Huge
   \textbf{\textbf{Supplementary Information}\\
   \Large {\setstretch{1.0}
Magnon spin transport in the van der Waals antiferromagnet CrPS\textsubscript{4} for non-collinear and collinear magnetization} }
\end{center}

\section{Interface spin conversion efficiency}
To evaluate the spin current through the CrPS\textsubscript{4} and estimate the magnon conductivity, we look at the nature of the Pt/CrPS\textsubscript{4} interface. The magnon spin injection and detection at the Pt/CrPS\textsubscript{4} interface can be expressed in terms of the current transfer efficiency $\eta$, which is the conversion efficiency of the charge current to spin current injected. In case the detector circuit is shorted, $\eta=R_{nl}/R_0$, where $R_0$ is the electrical resistance of the injector Pt strip\cite{cornelissen_magnon_2016}. $\eta$ decreases algebraically scaling $1/d$ for $d\ll\lambda$ (Ohmic), and exponentially for $d\gg\lambda$, as is given by equation 1 in the main text. The ratio of the spin accumulation in the injector and detector, $\eta_s=\mu^{det}_s/\mu^{inj}_s\gg\eta$, for the charge to spin conversion by the SHE is inefficient. 

Cornelissen \textit{et al.}\cite{cornelissen_magnon_2016} showed that for Pt on YIG, for $d\ll\lambda$ the non-local magnon signal, which is Ohmic for diffusive magnon transport, is sensitive to the interface. This would lead to a deviation (decrease in signal) from the $1/d$, Ohmic, dependence $R^{1\omega}_{nl}$ as a function of $d$ for small values of $d$. In our results we do not observed a clear deviation. Therefore, we assume the spin transport is dominated by the magnon transport properties, $\sigma_m$ and $\lambda$, of CrPS\textsubscript{4}.

Hence, to get the spin accumulation conversion coefficients, $\eta_{inj}=\mu_s/(eI)$, at the injector and detector we calculated the electrical spin accumulation, $\mu_s$ (expressed in energy), which is generated at the Pt/CrPS\textsubscript{4} interface:
\begin{equation}
    \mu_s=2eI\theta_{Pt}\frac{l_s}{\sigma_e tw}\tanh{\frac{t}{2l_s}},
    \label{eq:spinaccum.}
\end{equation}
where $e, t, w, \theta_{Pt}, l_s$ and $\sigma_s$ are the electron charge, the Pt film thickness and width, spin Hall angle, spin relaxation length, and conductivity of the Pt. With these parameters for Pt being $\theta_{Pt}$ = 0.11, $\sigma_e = 2\times10^6$ S/m, and $l_s = 1.5\times10^{-9}$ m, we calculated $\eta_{inj}$ = 0.077 $\Omega$ for Pt contacts with $w = 300$ nm.

\section{Extraction of the magnon conductivity}\label{sec:sigma_m}
By reciprocity, the conversion efficiency of the detector is $\eta_{det} = \eta_{inj}$. The magnon conductance, $G_m$, follows from $R_{nl}$, with $L$ being the length of the Pt strip:
\begin{equation}
    G_m = \frac{1}{\eta_{inj}\eta_{det}}\frac{V_{nl}}{I}=\frac{R_{nl}L}{\eta_{inj}\eta_{det}}.
    \label{eq:MagnonConductance}
\end{equation}
$\sigma_m$ can be given as a function of the thickness $t_{CPS}$ of the CrPS\textsubscript{4}. 
\begin{equation}
    \sigma_m = \frac{G_md}{t_{CPS}L}.
\end{equation}
For $t_{CPS}\ll\lambda_m$ and $t_{CPS}\ll L$, as is the case for all our devices, we consider $\mu_m$ to be constant in the $z$-direction. Equation 1, in the main text, can then be rewritten\cite{takahashi_spin_2003,wei_giant_2022}:
\begin{align}
    R_{nl}=\frac{\sigma_m t_{CPS}\eta_{inj}\eta_{det}}{\lambda_m}\text{csch}\frac{d}{\lambda_m}\rightarrow \nonumber\\ 
    \begin{cases}
      \frac{\sigma_m t_{CPS}\eta_{inj}\eta_{det}}{d} & \text{for } d\ll\lambda_m\\
      \frac{2\sigma_m t_{CPS}\eta_{inj}\eta_{det}}{\lambda_m}\exp{\left(-\frac{d}{\lambda_m}\right)} & \text{for } d\gg\lambda_m\\ 
    \end{cases}
    \label{eq:magnonconductivityNL}
\end{align}
From equation 1, in the main text, and \ref{eq:magnonconductivityNL} we can deduce that when $d$ is smaller than $\lambda_m$ transport is in the Ohmic regime (only magnon diffusion) and the magnons are conserved. When $d$ is larger than $\lambda_m$, the non-local signal decays exponentially as a function of $d$, since the number of magnons decays\cite{takahashi_spin_2003, cornelissen_long-distance_2015,wei_giant_2022}. For the Ohmic regime, we can then derive the following equation for $\sigma_m$:
\begin{equation}
    \sigma_m = -\frac{C}{2t_{CPS}\eta_{inj}\eta_{det}},
    \label{eq:sigma_m_C}
\end{equation}
which we use to extract $\sigma_m$ in figure 5 in the main text.

\section{Devices in this manuscript}
All experimental data discussed in this manuscript is measured on three different non-local magnon transport devices of 7 nm thick Pt strips on top of exfoliated CrPS\textsubscript{4} flakes of varying thicknesses. The silicon (Si\textsuperscript{++}) with a pre-grown 285 nm thick SiO\textsubscript{2} layer is used as a substrate. If not specifically indicated, all results discussed in this manuscript are measured on device D1. An optical image of this device is shown in figure 1 in the main text. The CrPS\textsubscript{4} thickness of this device is $\sim$160 nm thick. Device D2 and D3, are very similar to D1 and have a thickness of $\sim$100 nm. In our measurements we did not observe a dependence of the measured signal as a function of CrPS\textsubscript{4} thickness. We expect only for few layers of CrPS\textsubscript{4} (thickness $<$$\sim$50 nm) the thickness will play an important role, as is observed in Pt/YIG\cite{wei_giant_2022}.
\begin{figure}[b!]
    \centering
    \includegraphics[width=0.9\linewidth]{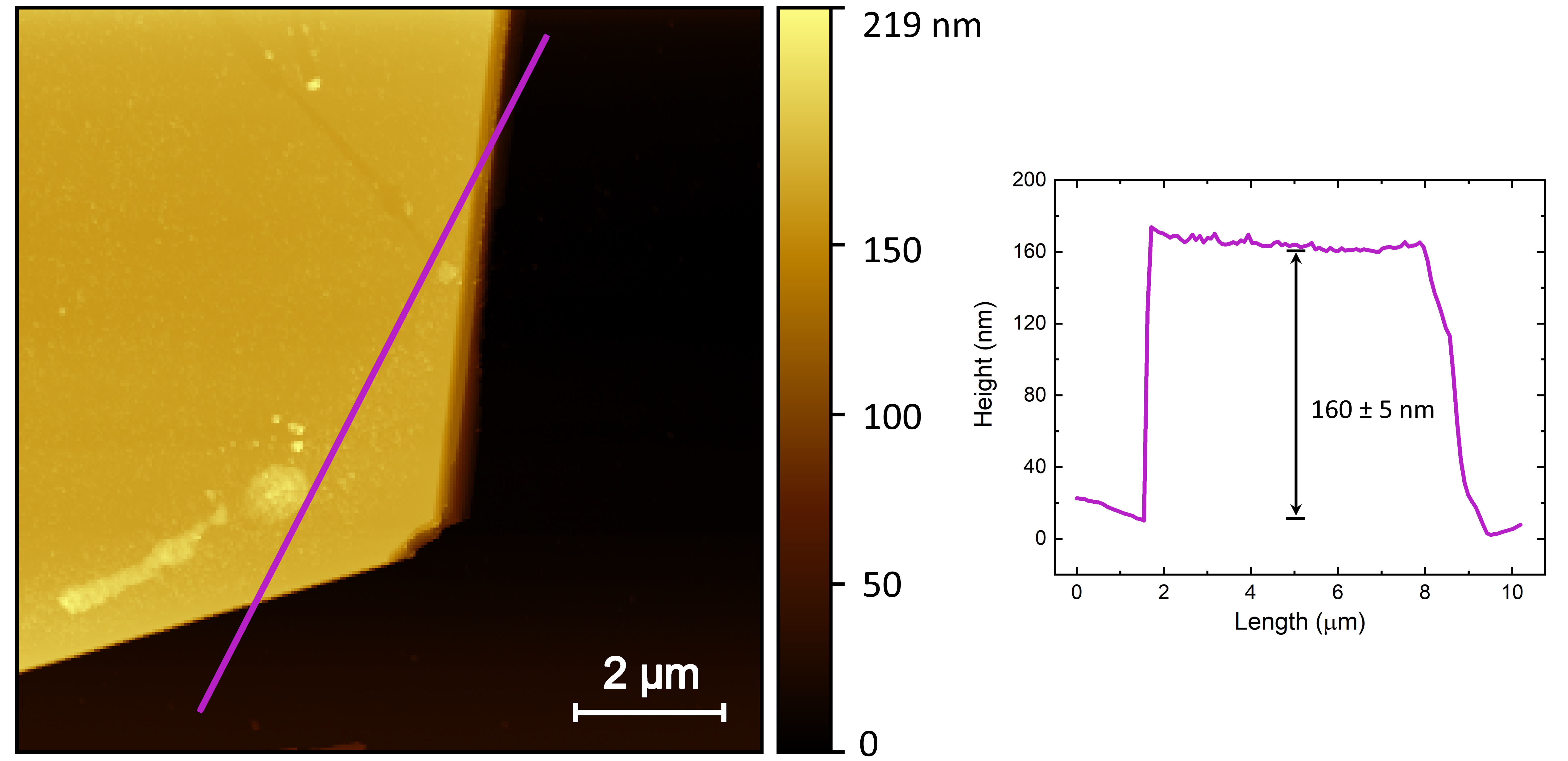}
    \caption{AFM image of device D1 with a line profile to determine the flake thickness on the right of the figure.}
    \label{fig:AFM}
\end{figure}

This thickness of the CrPS\textsubscript{4} flakes is determined using atomic force microscopy (AFM) using a Bruker Multimode AFM. An AFM scan for device D1 is given in figure \ref{fig:AFM}. The thicknesses of the other devices and the Pt strips are determined in similar fashion.

\section{Magnetrometry measurements CrPS\textsubscript{4}}
\begin{figure}[h]
    \centering
    \includegraphics[width=0.95\linewidth]{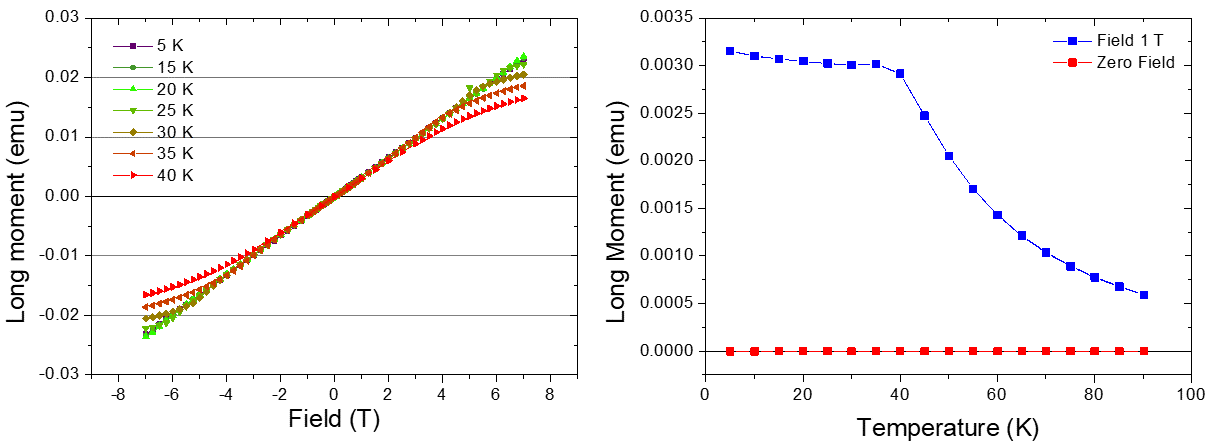}
    \caption{Magnetization behavior for a bulk crystal of CrPS\textsubscript{4}, with external field applied along the b-axis. \textit{Left}: Magnetization as a function of  appied field. At a field of 7T, at temperatures $>20$K the magnetization start to saturate. For large temperatures the saturation start at lower external fields. \textit{Right}: Magnetization as a function of temperatures at zero applied field and at 1T. Data is obtained using a Quantum Design MPMS magnetometer.}
    \label{fig:S1}
\end{figure}

Bulk CrPS\textsubscript{4} crystals of $\sim0.5$mm thick are probed using a Quantum Design MPMS-XL magnetometer. The response of these bulk samples to an applied magnetic field along the b-axis (in-plane) of the crystal is given in figure \ref{fig:S1} as a function of field and temperature. The magnetization starts to saturate around 7 T for temperatures $>$20 K. Furthermore, we do not observe a spontaneous magnetization at zero applied field, which is expected from an antiferromagnet. At an external field of 1 T, we find the Néel temperature is around 40 K. Note that we measure a non-zero magnetization here as the finite field is canting the two magnetic sublattices, yielding a small net magnetization even at temperatures slightly above the Néel temperature. This magnetization data agrees well with the measurements reported by Peng \textit{et al.}\cite{peng_magnetic_2020}.

\begin{figure}[b!]
    \centering
    \includegraphics[width=0.95\linewidth]{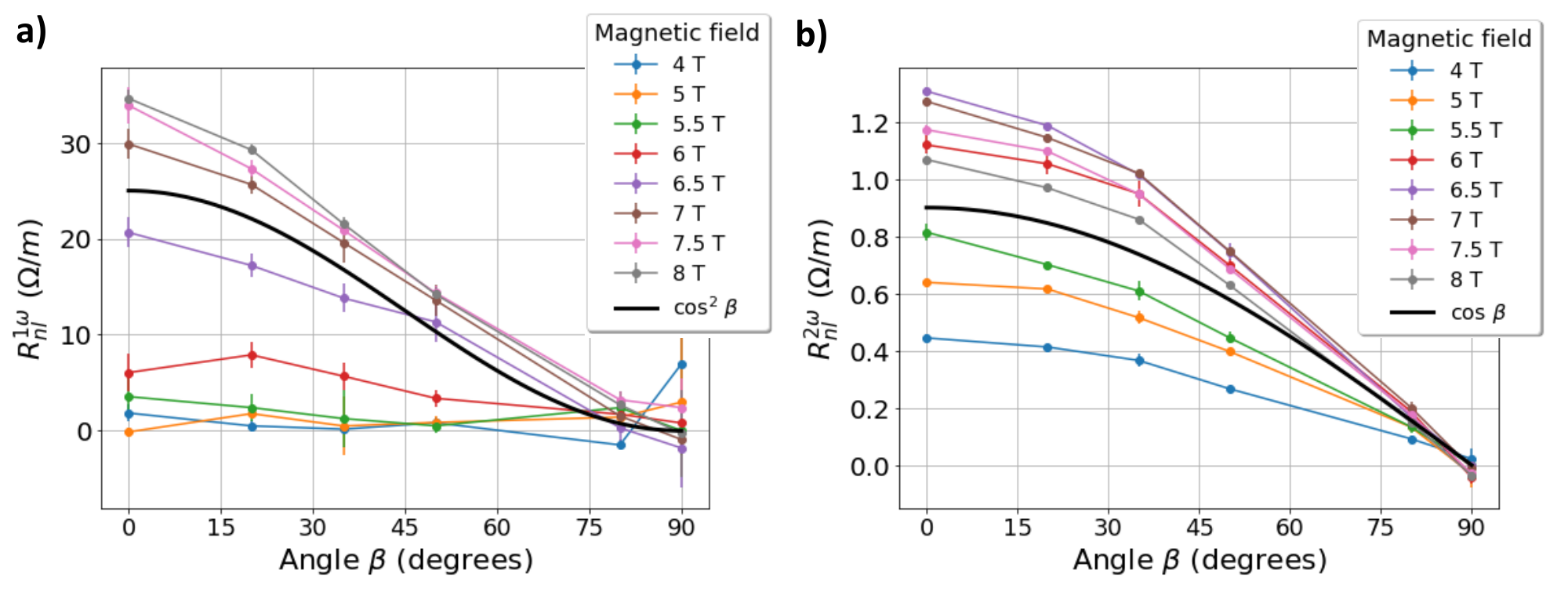}
    \caption{$R^{1\omega}_{nl}$ (a) and $R^{2\omega}_{nl}$ (b) as a function of out of plane angle $\beta$ at different field strengths, T = 25K, $I_{AC}=80\mu$A and $d=270$nm. The black line corresponds to a $\cos^2{\beta}$ or $\cos{\beta}$ of arbitrary amplitude and is only for comparison purposes.}
    \label{fig:oop_beta}
\end{figure}

\section{Out of plane angle ($\beta$) measurements}
In figure \ref{fig:oop_beta} the non-local resistances $R^{1\omega}_{nl}$ and $R^{2\omega}_{nl}$ are given for  applied fields at angle $\beta$ with respect to the CrPS\textsubscript{4} crystal c-axis (oop), note that $\beta=90^{\circ}-\theta_{\perp}$. $R^{1\omega}_{nl}$ and $R^{2\omega}_{nl}$ are expected to follow the projection of the net magnetization on the y-direction (see figure 1 in the main text), i.e. $\cos^2{\beta}$ and $\cos{\beta}$, respectively. As only magnon spin with a component polarized along the y-direction can be picked up or injected by the Pt contacts. The black line in figure \ref{fig:oop_beta} represents a $\cos^2{\beta}$ (Fig. \ref{fig:oop_beta}a) or $\cos{\beta}$ (Fig. \ref{fig:oop_beta}b) of arbitrary amplitude for comparison purposes. 

\begin{figure}[b!]
    \centering
    \includegraphics[width=0.90\linewidth]{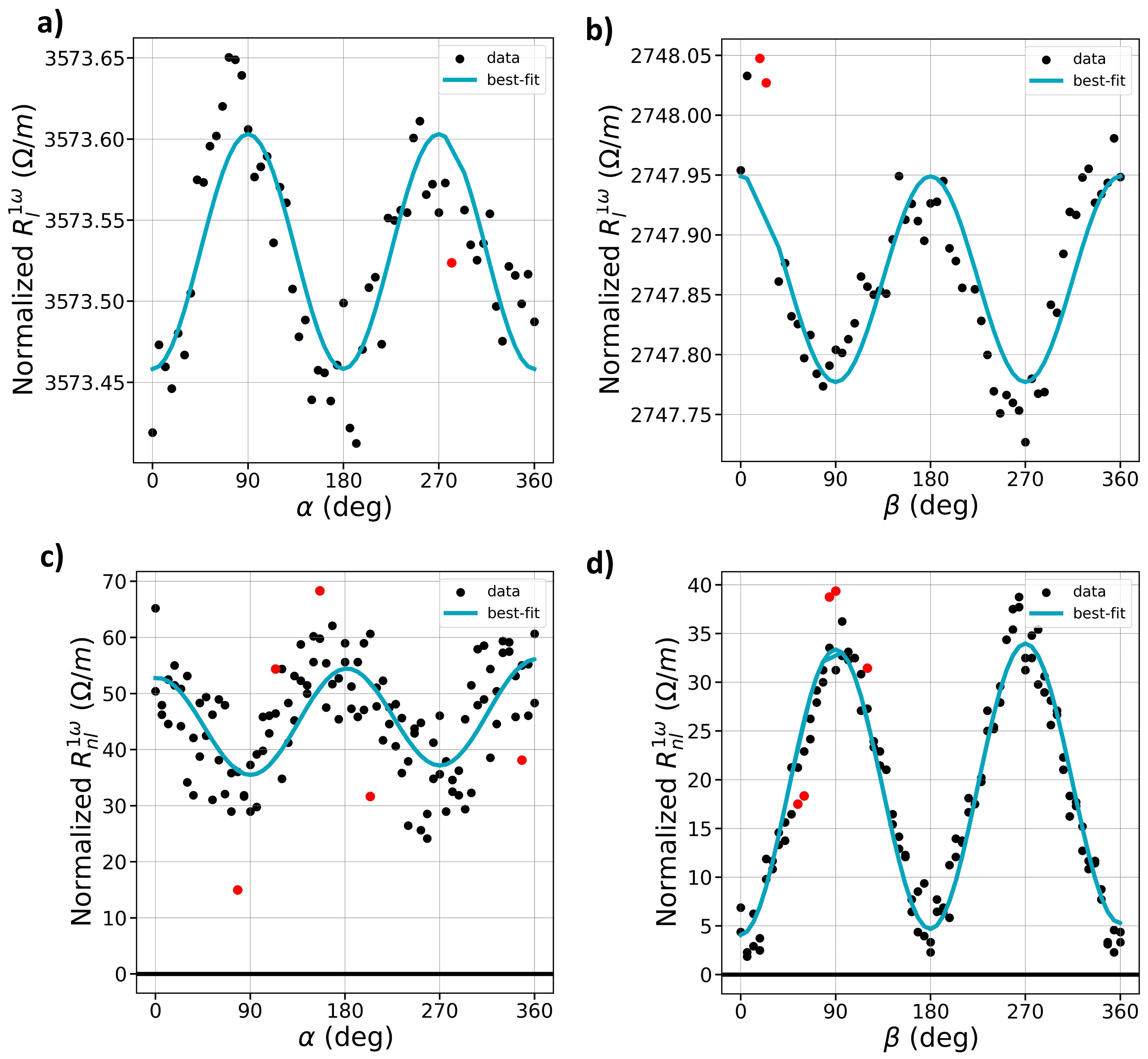}
    \caption{Non-local ADMR measurements at $T=25$K $H=7.5$T for device D2 for in plane ($\alpha$) and Device D1 out of plane to in plane perpendicular to the Pt strip ($\beta$) for two Pt strips with edge-to-edge spacing of 360 nm (a,c) and 270 nm (b,d). Here $\alpha=0$ when the field is applied perpendicular to the Pt strip for which $\boldsymbol{\mu}\parallel \mathbf{M}$ (where $\mathbf{M}$ is the net magnetization of CrPS\textsubscript{4} and $\alpha=90^{\circ}$ and $\beta=0$ for a field fully out of plane and $\beta=90^{\circ}$ equals $\alpha=0$. No offset has been removed for any of these plots. The in plane and out of plane measurements are performed in different cryostat inserts, resulting in different offsets for the in plane and out of plane measurements. The blue fitted line, is a guide to the eye, no values were extracted from these particular fits. }
    \label{fig:S2}
\end{figure}

\subsection*{Angular dependent magnetoresistance measurements}
Besides measurements of the non-local signals as a function of field at fixed angles, also angular dependent magnetoresistance (ADMR) measurements are performed, both for in plane and out of plane fields. In figure \ref{fig:S2} both the local resistance ($R^{1\omega}_{l}$) modulation (a,b), caused by the spin Hall magnetoresistance (SMR) as well and the non-local resistance ($R^{1\omega}_{nl}$) (c,d) are given. The angular dependences of $R^{1\omega}_{l}$ and $R^{1\omega}_{nl}$ as a function of $\alpha$ are given by:
\begin{align}
    R^{1\omega}_{l}     &= R^{1\omega}_{l,0} + \Delta R^{1\omega}_{l}\sin^2{\alpha}, \\
    R^{1\omega}_{nl}    &= \Delta R^{1\omega}_{nl}\cos^2{\alpha} + \text{offset},\label{eq:RNL_Alpha}
\end{align}
and as a function of $\beta$:
\begin{align}
    R^{1\omega}_{l}     &= R^{1\omega}_{l,0} + \Delta R^{1\omega}_{l}\cos^2{\beta}, \\
    R^{1\omega}_{nl}    &= \Delta R^{1\omega}_{nl}\sin^2{\beta} + \text{offset}.\label{eq:RNL_Beta}
\end{align}
Here $R^{1\omega}_{l,0}$ is the initial resistance of the Pt strip and the resistances introduce by the sample cryostat insert and $\Delta R^{1\omega}_{l/(nl)}$ is the local (non-local) resistance modulation amplitude. Using these equations to fit to the ADMR measured data, these parameters can be extracted. This procedure is use to extract the $\Delta R^{1\omega}_{nl}$ for devices D2 and D3 which are used in figure 5 in the main text.
\begin{figure}
    \centering
    \includegraphics[width=\linewidth]{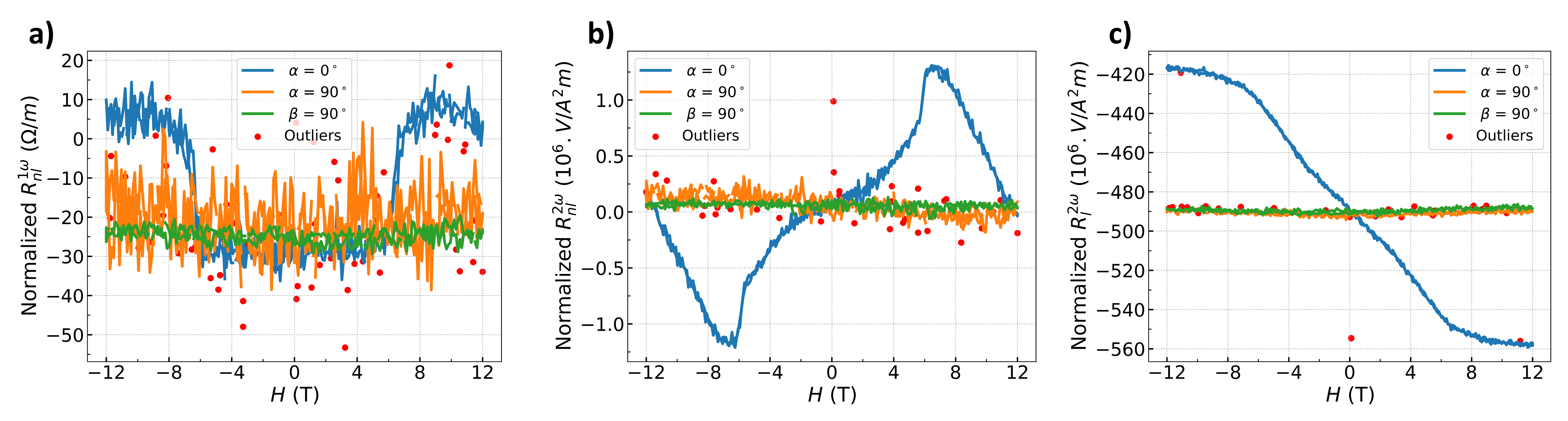}
    \caption{$R^{1\omega}_{nl}$, $R^{2\omega}_{nl}$ and $R^{2\omega}_{l}$ (left to right) as a function of external field at $\alpha=0$ (in plane perpendicular to the Pt strip),  $\alpha=90^{\circ}$ (in plane parallel to the Pt strip) and $\beta=90^{\circ}$ (out of plane). Outliers are marked in red. $d=270$nm, $T=25$K and $I_{AC}=80\mu$A. }
    \label{fig:S3}
\end{figure}

The $\Delta R^{1\omega}_{nl}$ extracted from device D1 used in the main text are all extracted from the field dependences of $R^{1\omega}_{nl}$ as given in figure 3 in the main text. In order to determine the `offset' value in $R^{1\omega}_{nl}$ (see equation \ref{eq:RNL_Alpha} and \ref{eq:RNL_Beta}), such that we can find $\Delta R^{1\omega}_{nl}$, also a field scan direction of $\gamma$ (which is in the $\alpha=90^{\circ}$ and $\beta=0$ plane) is performed and is given in figure \ref{fig:S3}.

Only for $\alpha=0$, $R^{1\omega}_{nl}$, $R^{2\omega}_{nl}$ and $R^{2\omega}_{l}$ are non zero. Which corresponds to $\mathbf{H}\parallel\boldsymbol{\mu}$. Note that there is a 5$^{\circ}$ offset in both $\alpha$ and $\beta$, leading to a small deviation from the exact $\alpha=90^{\circ}$ and $\beta=90^{\circ}$. The clear onset around the spin-flip field and saturation at larger fields for $\alpha=0$ in figure \ref{fig:S3}a is taken as the non-local resistance. The value of $R^{1\omega}_{nl}$ in between -4T and 4T is taken as the offset.

\newpage
\section{Magnon conductivity $\sigma_m$ at $H>8$T fields}
\begin{figure}[h]
    \centering
    \includegraphics[width=\linewidth]{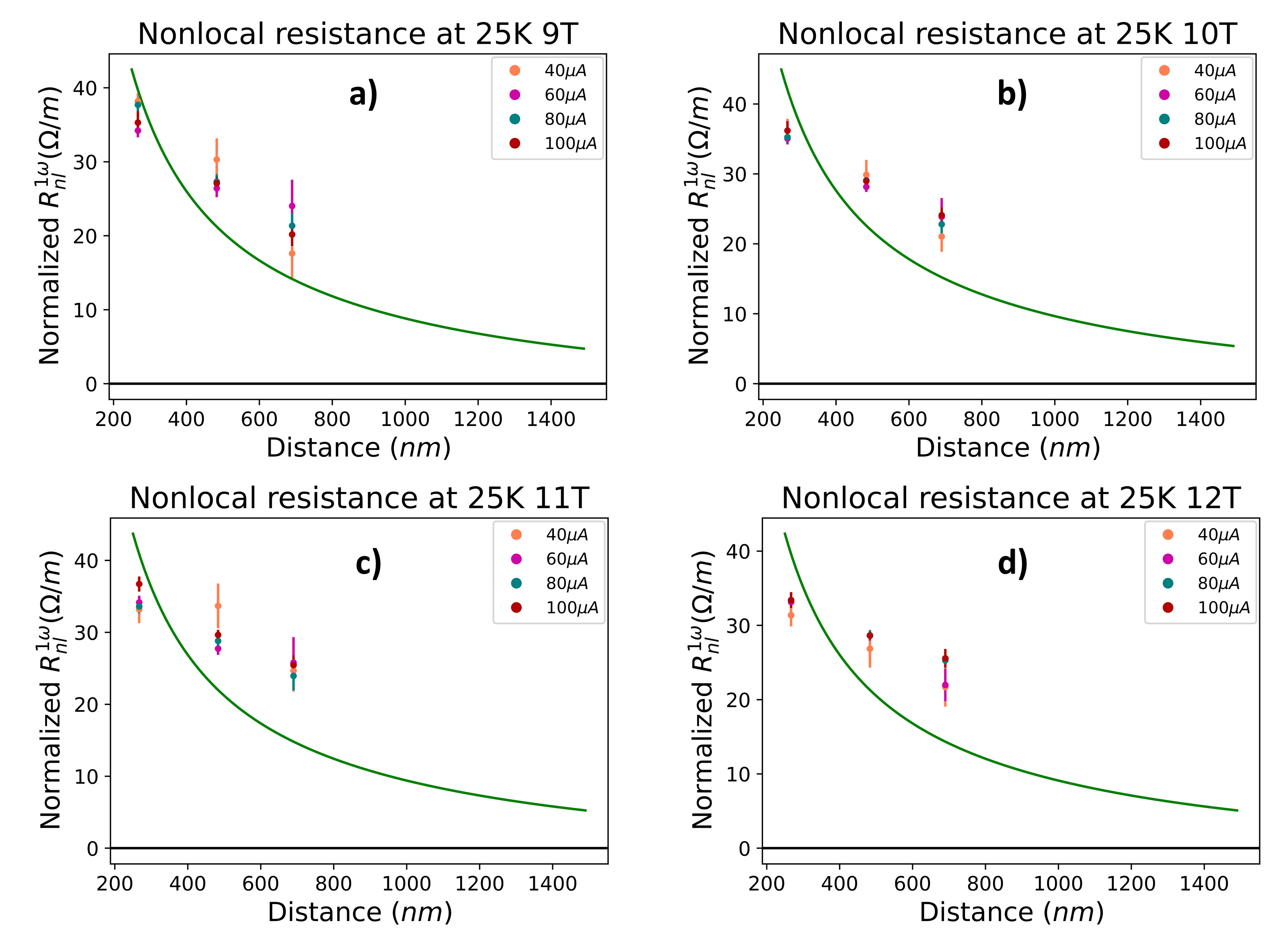}
    \caption{$R^{1\omega}_{nl}$ at $H>8$T. These measurements are only from device D1. The green line is the fit result using equation 1 as fit function.}
    \label{fig:S4}
\end{figure}

In figure 5b in the main text, the magnon conductivity $\sigma_m$ is given as a function of in-plane field strength. The values are extracted as is explained in section \ref{sec:sigma_m} in this supplementary information. For fields $H>8$T, the dependence of $R^{1\omega}_{nl}$ on the distance between the injector and detector Pt strip does not follow the relation given in equation 1 in the main text. In figure \ref{fig:S4}, the measured $R^{1\omega}_{nl}$ for $H>8$T are given. Under the assumption that $\lambda_m>800$nm, which is the relaxation length at 8T, magnon transport is in the purely diffusive (Ohmic) regime and we therefore would expect a 1/$d$ decrease in $R^{1\omega}_{nl}$. This  is not the case, as can be seen in figure \ref{fig:S4}. We cannot explain this behavior of magnon transport. However, it is clear that $R^{1\omega}_{nl}$ does saturate at these fields. We therefore extract $\sigma_m$ directly from the values of $R^{1\omega}_{nl}$ in figure \ref{fig:S4}. The $\sigma_m$ given in figure 5b in the main text, are taken for $d=480$nm and are calculated by equation \ref{eq:magnonconductivityNL} for $d\ll\lambda_m$ for $t_{CPS}=160$nm. 

\newpage
\section{Temperature dependence of $R^{2\omega}_{nl}$ and $R^{2\omega}_{l}$}
\begin{figure}[h]
    \centering
    \includegraphics[width=\linewidth]{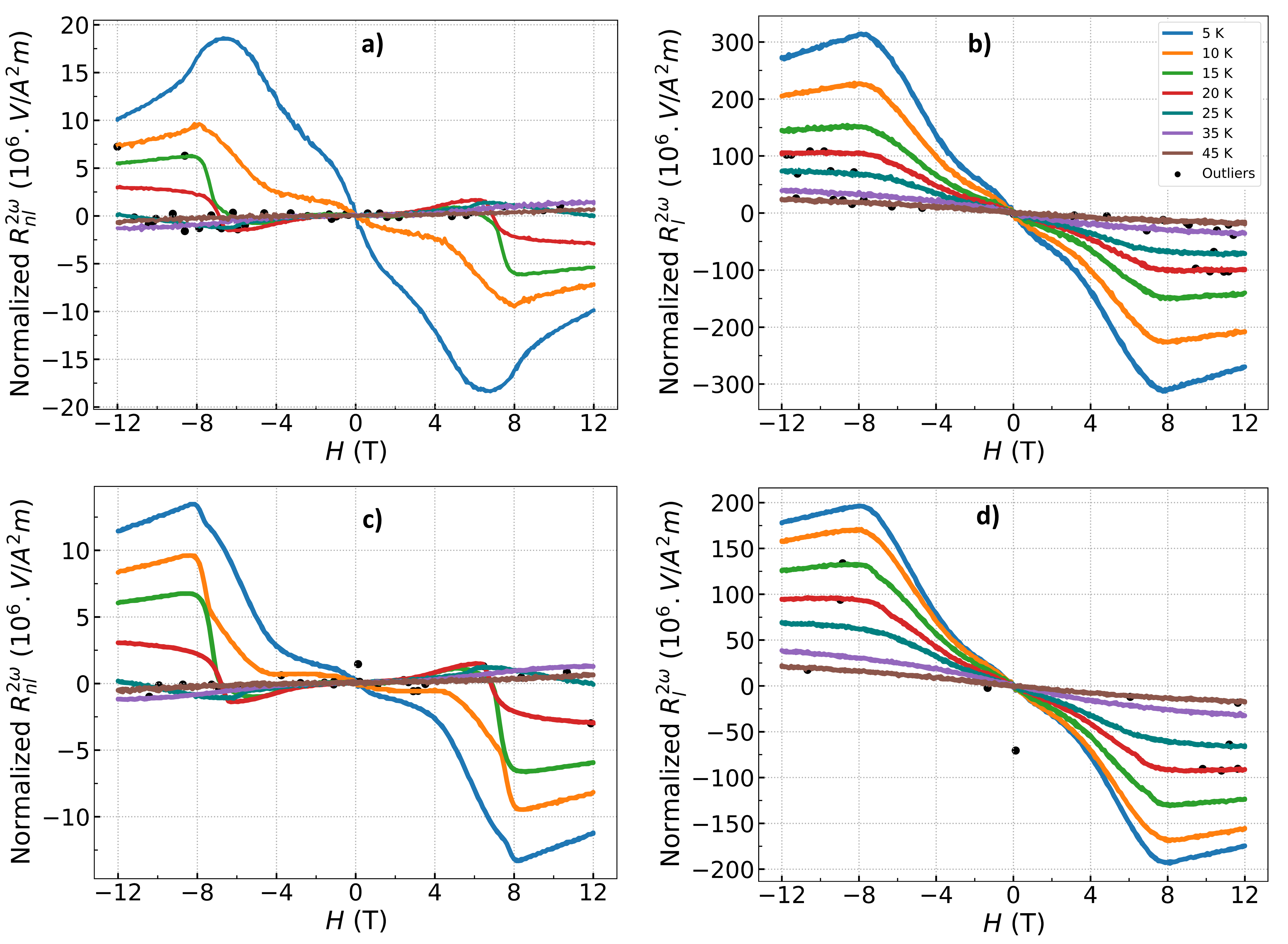}
    \caption{Temperature dependence of $R_{nl}^{2\omega}$ and $R_{l}^{2\omega}$ at (a), (b) $I=60\mu$A and (c), (d) $I=100\mu$A for $d=270$ nm. Data points in black are outliers. For $R_{l}^{2\omega}$ in (b) and (d), a constant offset is removed.}
    \label{fig:S5}
\end{figure}

In figure \ref{fig:S5}, $R_{nl}^{2\omega}$ and $R_{l}^{2\omega}$ are shown as a function of field for different temperatures. The strong increase of $R_{nl}^{2\omega}$ (Fig. \ref{fig:S5}a and \ref{fig:S5}c) above the spin-flip field and the sign change at larger temperature are not completely understood.

\newpage


\end{document}